\documentclass[%
 reprint,
 nofootinbib,
 amsmath,amssymb,
 pra,
 twocolumn,
 notitlepage
]{revtex4-1}

\usepackage{amsmath,amssymb,amsfonts}
\usepackage{algorithm}
\usepackage{algpseudocode}
\usepackage{algorithmicx}
\usepackage{graphicx}
\usepackage[font=footnotesize,labelfont=bf,justification=raggedright]{caption}

\usepackage{textcomp}
\usepackage{braket}
\usepackage{bm}
\usepackage{color}
\def\BibTeX{{\rm B\kern-.05em{\sc i\kern-.025em b}\kern-.08em
    T\kern-.1667em\lower.7ex\hbox{E}\kern-.125emX}}

\usepackage{hyperref}
\newcommand{\boldvec}[1]{\overrightarrow{\text{\textbf{#1}}}}

\begin{document}

\title{Efficient Quantum Circuits for Accurate State Preparation\\of Smooth, Differentiable Functions}

\author{Adam Holmes}
\email{adholmes@uchicago.edu}
\affiliation{The University of Chicago, Chicago, IL 60615, USA}
\affiliation{Intel Labs, Hillsboro, OR 97124, USA}

\author{A. Y. Matsuura}
\affiliation{Intel Labs, Hillsboro, OR 97124, USA}

\date{\today}

\begin{abstract}
Effective quantum computation relies upon making good use of the exponential information capacity of a quantum machine. A large barrier to designing quantum algorithms for execution on real quantum machines is that, in general, it is intractably difficult to construct an arbitrary quantum state to high precision. Many quantum algorithms rely instead upon initializing the machine in a simple state, and evolving the state through an efficient (i.e. at most polynomial-depth) quantum algorithm. In this work, we show that there exist families of quantum states that can be prepared to high precision with circuits of linear size and depth. We focus on real-valued, smooth, differentiable functions with bounded derivatives on a domain of interest, exemplified by commonly used probability distributions. We further develop an algorithm that requires only linear classical computation time to generate accurate linear-depth circuits to prepare these states, and apply this to well-known and heavily-utilized functions including Gaussian and lognormal distributions. Our procedure rests upon the quantum state representation tool known as the \textit{matrix product state} (MPS). By efficiently and scalably encoding an explicit amplitude function into an MPS, a high fidelity, linear-depth circuit can directly be generated. These results enable the execution of many quantum algorithms that, aside from initialization, are otherwise depth-efficient. 
\vspace{-0.5cm}
\end{abstract}

\maketitle

\section{Introduction}
\label{sec:introduction}
%

Many applications depend upon initialization of a quantum register into specific states. A good example of this is the class of Monte Carlo style quantum algorithms that can compute expectation values of functions over classical probability distributions with quadratic speedup over all-classical counterparts \cite{stamatopoulos2019option,martin2019towards,woerner2019quantum,montanaro2015quantum,harrow2009quantum}. Other examples include machine learning classical training data sets that are used as input to a quantum machine learning algorithm \cite{lloyd2014quantum,rebentrost2014quantum,harrow2009quantum,wiebe2012quantum,childs2017quantum}. Preparing the state corresponding to the data in both of these cases is in general an exponentially hard problem \cite{plesch2011quantum} that is often omitted from application performance analysis. However, if initialization is this difficult these applications may not be viable for quantum computing. To address this issue, we develop a technique for generating circuits that can prepare certain families of quantum states efficiently. These families of states may enable the execution of quantum algorithms  that depend on classical input data on smaller, near-term machines.


We focus on smooth, differentiable, real-valued (SDR) functions with bounded derivatives, and show that they can be constructed efficiently with linear-depth quantum circuits. The primary contribution of this work is Algorithm \ref{alg:main_algorithm}, that combines a novel encoding of piecewise polynomial function approximators into matrix product state form, with existing techniques for MPS compression and low-rank MPS quantum gate extraction. The algorithm can be tuned, and different configurations have different impacts on error and scalability. 

Sections \ref{sec:related} and  \ref{sec:background} discuss related work and set the notation and background necessary to understand the techniques that build our algorithm. Section \ref{sec:theory} shows theoretically that these types of functions display important properties, among these an exponentially decreasing von-Neumann entropy as qubits are added to the states. As a result, they have efficient MPS representations with low bond-dimension. 

Section \ref{sec:techniques} describes our techniques, including explicit piecewise polynomial MPS function approximation, MPS arithmetic, and variational MPS approximation with low, bounded bond-dimension. Leveraging an algorithm developed in \cite{ran2020encoding} that constructs a linear-depth quantum circuit from a MPS, we show that we can construct quantum circuits that prepare our desired amplitude functions with a high degree of accuracy.

Section \ref{sec:algorithm} presents our algorithm for efficiently and scalably constructing the state preparation circuit corresponding to the functions we study. The algorithm combines techniques described in \ref{sec:techniques}, first approximating the target function as a piecewise polynomial, encoding each piece into an MPS, variationally compressing the MPS into one of low-rank, and extracting gates directly from this resulting state. This combination is computationally tractable, requiring $\mathcal{O}(N)$ computation, and is bottlenecked by the variational MPS compression. Stages of the algorithm can be modified as desired, for example modifying the functional form of each component of the function approximation. The result is a tunable algorithm of linear complexity in the size of the system that prepares approximated analytical quantum states with linear size and depth circuits.

Section \ref{sec:results} analyzes the performance of our algorithm targeting representative examples of SDR functions -- namely Gaussian, lognormal, and Lorentzian distributions. We show numerical analysis of the accuracy of our circuits for small system sizes, and demonstrate that the techniques can prepare states with good accuracy across a range of function parameter regions. 

Section \ref{sec:applications} discusses how these results can be used to enable classes of quantum algorithms designed to estimate the expectation value of linear functions over probability distributions, and Section \ref{sec:conclusion} concludes.
\section{Related Work}
\label{sec:related}
While several techniques for state preparation have been proposed, they are often expensive in either the classical compute requirements of the algorithm or the resulting quantum gate sequence. Three examples are: exact Schmidt-Decomposition \cite{coles2018quantum}, recursive integrable distribution encoding \cite{grover2002creating}, and quantum Generative Adversarial Networks (qGAN) \cite{zoufal2019quantum}. 

The most general techniques for state preparation rely primarily on iterative Schmidt-Decomposition algorithms that require exponential classical computation, scaling that is prohibitive for large states \cite{coles2018quantum}. Early work from Grover \cite{grover2002creating} utilized a recursion of integral calculations that ultimately requires an exponential number of angles and classical calculations, even though the result is a potentially efficient quantum circuit \cite{garcia2019quantum}. Others have proposed qGAN learning-based approaches \cite{zoufal2019quantum} that rely on a classically expensive combination of a variational quantum
circuit and a classical neural network. These techniques construct $\mathcal{O}(poly(N))$ sized quantum circuits corresponding to the learned distributions, which have accuracy corresponding to the effectiveness of the overall learning technique. 

The fundamental difference between our work and these presented methods is that our algorithm can construct accurate \textit{linear size and depth} quantum circuits, and only requires \textit{linear} classical computation time to do so. 



\section{Background}
\label{sec:background}
This work develops a technique to construct quantum states in which each binary-indexed basis state corresponds to a coefficient that follows a specific amplitude function. We restrict our focus to smooth, differentiable, real-valued (SDR) functions that have bounded derivatives.

\subsection{Notation} 

In general, for some SDR function $f(x)$ with support over some domain $D$ we first discretize $D$ into $2^N$ points for a system of $N$ qubits, evenly placed across the domain. This uses a mapping from index $k$ to domain values $x(k)$:
\begin{equation}
    x(k) = a + \frac{kL}{h}
    \label{eq:index_domain_map}
\end{equation}
where $D = [a,b]$, $L = b-a$ is the width of the domain, and $h = 2^N-1$ is the number of gridpoints. Our goal is then to construct quantum states $\ket{\psi_{f(x)}}$ defined as:
\begin{equation}
\ket{\psi_{f(x)}} = \bigotimes_{k=0}^{N-1} \sqrt{f(x_{S_k})}\ket{S_k} 
\end{equation}
where we will use the big-endian binary encoding of a length $N$ binary string $S$ written as $s_0, s_1, \cdots, s_{N-1}$ with each individual bit $s_i \in \{0,1\}$. Big-endian here defines the mapping of binary strings to integers as:
\begin{equation}
    k = \sum_{i=0}^{N-1} s_i \times 2^{N-1-i} 
    \label{eq:binary_encoding}
\end{equation}
In this notation, we have that the function $f(x_{S_k})$ is the evaluation of the target SDR function $f(x)$ evaluated at the domain value $x(k)$ defined by the \textit{index} $k$ induced by the binary string $S_k$ from equation \ref{eq:binary_encoding}. Written together:
\begin{equation}
    \ket{\psi_{f(x)}} = \bigotimes_{k=0}^{N-1} \sqrt{f\bigg(a_0+L\frac{\sum_{i=0}^{N-1}S^k_{i}\times 2^{N-1-i}}{2^N}\bigg)} \ket{S_k}
    \label{eq:target_state}
\end{equation}
\subsection{Smooth, differentiable functions}
We are focusing in this work on smooth, differentiable, real-valued (SDR) functions, as these admit many properties that allow for their efficient construction inside a quantum machine. These functions have two properties that we will make use of:
\begin{itemize}
    \item discretization accuracy increases exponentially with the number of qubits included in the system, and
    \item the maximum Von-Neumann entropy of the state grows much more slowly as qubits are added.
\end{itemize}
Because of the two properties derived explicitly in \cite{garcia2019quantum}, we find a very useful scaling relationship: as the system scales up in qubit count, these functions admit efficient and accurate representations in their low-rank approximations, while the accuracy of the encoded state continues to exponentially increase.

Examples of these types of functions include probability distributions, particularly Gaussian, lognormal, and Lorentzian distributions. We will see that the accuracy of our techniques is dependent on the smoothness of these distributions, specifically with the upper bound on the derivative of the distribution on the domain of interest: $\Tilde{f'} = \max_{k} |f'(x_k)|$ for $k \in \{0,2^{N}-1\}$. For distributions that are relatively slowly changing, $\Tilde{f'}$ is small, which leads to a more exact representation of the discretized function in a low-rank approximation.

This relationship encourages the use of efficient representations of low-rank matrix approximations, and one that is particularly suited to this task is the \textit{matrix product state}.
\subsection{Matrix product states}
Existing literature surrounding these mathematical techniques is vast within the physics community \cite{perez2006matrix,schollwock2011density,zaletel2015time} as well as within computational mathematics and scientific computing \cite{oseledets2011tensor,holtz2012alternating,lee2015estimating} where these are referred to as tensor trains. Here we describe only properties relevant to this work.

A \textit{matrix product state} (MPS) is a collection of $N$ tensors $M_i$ where, in general, tensors can be any collection of values $A \in \mathbb{R}^{I_1, I_2, \cdots, I_K}$ containing $K$ distinct indices, referred to as $K$-th order tensors. Scalars, vectors, and matrices are considered $0^{\text{-th}}$, $1^{\text{st}}$, and $2^{\text{nd}}$ order tensors respectively. Within a matrix product state, we restrict ourselves to 3rd-order tensors $M_i \in \mathbb{R}^{\alpha_{i-1},d,\alpha_{i}}$, where $d$ is the number of allowed levels in our qubit model -- here considered to be held at $d=2$. Each $M_i$ will be referred to as a \textit{core} of the MPS. The $\alpha_j$ are known as \textit{bond dimensions} \cite{schollwock2011density} or \textit{virtual dimensions} that connect the matrices together. The maximum bond dimension is denoted as $\chi = \max_i \alpha_i$, a figure can be used to upper bound the computational complexity of many routines involving the MPS. In this work, we will write each of these tensors as $M_{\alpha_{i-1},\alpha_i}^{[i],s_i}$ to highlight the connection between the $i$-th tensor and the binary digit $s_i$ contained in an indexing string. The MPS $M_{\psi}$ can then be written as:
\begin{align}
    M_{\psi} &= \sum_{\bm{\alpha}, \bm{S}} M_{\alpha_1,\alpha_2}^{[1],s_1}M_{\alpha_2,\alpha_3}^{[2],s_2}...M_{\alpha_{N},\alpha_{N+1}}^{[N],s_{N}}
    \label{eq:MPS_explicit}
\end{align}
where the outer summations run over all bond dimensions $\bm{\alpha} = \alpha_1, \alpha_2, \cdots, \alpha_{N+1}$ and over all $2^N$ binary strings $\bm{S} = S_1, S_2, \cdots, S_{2^N}$. Left and right boundary bond dimensions $\alpha_1, \alpha_{N+1}$ are assumed to be equal to 1.

The efficiency of this representation can be seen by looking at it as an expression of $2^N$ values by writing them as a matrix product. In so doing, we only need to store $\mathcal{O}(2N\chi^2)$ values, 
While the MPS representation is capable of representing exactly any $2^N$ dimensional vector by allowing $\chi$ to grow unbounded, in our study of SDR functions we will find that holding $\chi = 2$ allows for highly accurate, compact representation of SDR functions, while keeping memory cost to a modest $\mathcal{O}(N)$.

\subsection{Tensor networks}
\begin{figure}[t!]
  \centering
  \begin{tabular}[b]{c}
    \includegraphics[width=.5\linewidth,trim={0 0 0 0},clip]{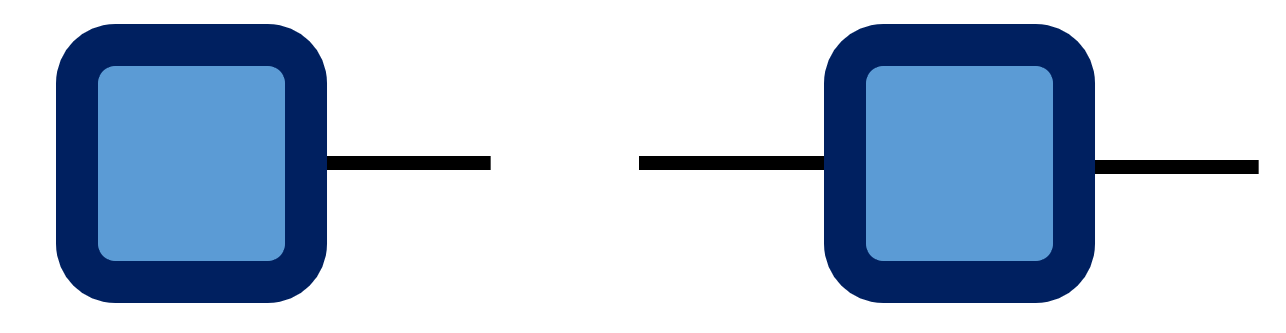} 
        \label{fig:vector_matrix}\\
   \end{tabular} \qquad
  \begin{tabular}[b]{c}
    \includegraphics[width=.5\linewidth,trim={0 0 0cm 0},clip]{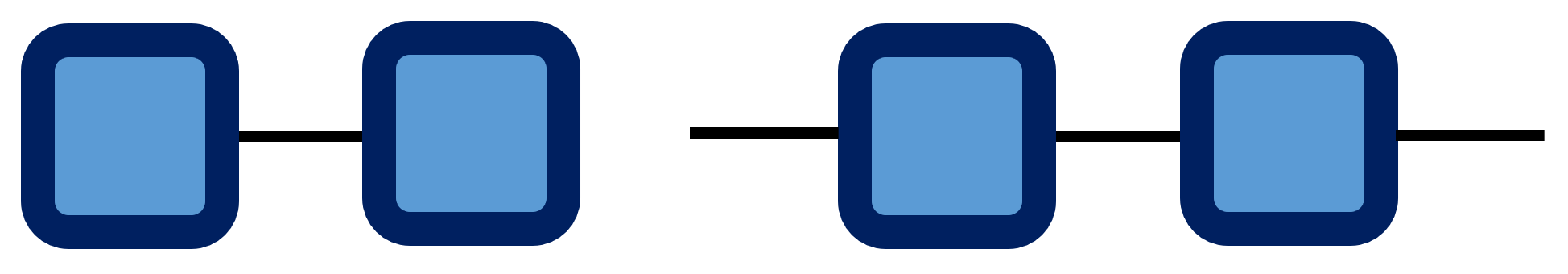} 
        \label{fig:vector_matrix_product}\\
    \end{tabular}
  \caption{(top) Left: vector | Right: matrix. \\(bottom) Left: inner product | Right: matrix product}
  \label{fig:graphical_tensors}
  \vspace{-0.5cm}
\end{figure}
MPS representations are just one among a family of these types of representations, where the order of all involved tensors can change. In general, these types of representations come with very useful graphical representations, and are described extensively in literature as \emph{tensor networks} \cite{biamonte2017tensor}. 

Tensors can be depicted as nodes in a graph with edges representing tensor indices. Figure \ref{fig:graphical_tensors} depicts single index vectors and two-index matrices in this fashion. This technique also allows us to express \emph{tensor contractions}, which are generalizations of vector and matrix operations. Tensor contractions are operations in which one or more indices of two or more tensors are explicitly summed over, and removed from the network. As an example, the vector-vector inner product can be written as a tensor contraction of a single index,
as can the matrix-matrix product.

Graphically, we describe contractions of indices by connecting the corresponding index-edge of the network together. The top of Figure \ref{fig:graphical_tensors} depicts the vector-vector inner product, and the bottom the matrix-matrix product. The matrix-matrix graphically can be understood by the labelling of the indices from left to right as: $i, j, j, k$, including two $j$ indices for the right- and left- most indices of matrices $A$ and $B$, respectively, and summing over them. Naturally these concepts generalize to any number of indices, and are used to simplify tensor networks while controlling for the complexity of the multiplications.




\section{Theory}
\label{sec:theory}
SDR functions have two desirable properties: as qubits are added to the system, discretization accuracy increases exponentially, while von-Neumann entropy (VNE) and therefore entanglement grows much more slowly. This can be shown by analyzing first the discretization error of a uniform gridding of the domain $D$ on which an SDR $f(x)$ has support, followed by studying the VNE of the constructed state.

\subsection{Discretization error}
The discretization error of the encoding of an SDR function $f(x)$ into a uniform grid of $2^N$ points across a domain $D =[a,b] \subset \mathbb{R}$ scales as $\mathcal{O}(2^{-N})$. To see this, first let $\Tilde{f'} = \max_x |f'(x)|$ be the maximum value of the derivative for $x \in D$. For a uniform gridding of the domain, we have $2^N$ exact values on which $f(x)$ is evaluated. Error arises when $x$ values are sampled that lie between any two gridpoints $(x_k, x_k+h)$ with $h=\frac{|b-a|}{2^N-1}$, and the discretized function approximation $\hat{f}(x)$ must be interpolated. Letting the maximum evaluation error occur at $\Tilde{x} = x_k+\delta$ for $\delta < 2^{-N}$, it is well known that uniform gridding and forward-difference interpolation produces first-order error linear in the step size, which in this context scales inverse-exponentially with the number of qubits in our system: $\mathcal{O}(h) = \mathcal{O}(2^{-N})$ \cite{trench2013elementary}.

As a result, the discretization error asymptotically halves when moving from a system of $N$ to $N+1$ qubits, or equivalently the function approximation accuracy doubles.

\subsection{von-Neumann Entropy Bound}
The von-Neumann Entropy (VNE) change of a system induced by adding $j$ qubits can be defined as the maximum VNE achieved by any Schmidt-decomposition partitioning of the system into subsystems of sizes $j$ and $N$ \cite{nielsen2002quantum}, as:
\begin{align}
    \Delta\text{VNE}(\ket{\psi}) = \max_j \sum_k \big(\lambda_k \log \lambda_k\big)_j
\end{align}
where 
\begin{align*}
    \ket{\psi} = \sum_k \lambda_k^{1/2} \ket{\phi^{1\cdots j}} \otimes \ket{\phi^{j \cdots N}}
\end{align*}
is the $j$-th Schmidt decomposition partitioning of the system $\ket{\psi}$.

This was studied extensively in \cite{garcia2019quantum}, and a bound was derived for the entropy of adding a single extra qubit to a state as:
\begin{align}
    \Delta\text{VNE}(\ket{\psi}) \leq \frac{L \sqrt{\Tilde{f'}}}{2^{N/2-1}} = \mathcal{O}(2^{-N/2})
    \label{eq:sdr_entropy}
\end{align}
This bound on the added entropy contribution from growing a discretized SDR $f(x)$ representation, along with the reduction in the discretization error of the same order $\mathcal{O}(2^{-N})$, implies that these states scale very efficiently in their representation. One reason for this is based on the idea that VNE is a proxy for the amount of entanglement contained within a state \cite{johri2017entanglement,garcia2019quantum,nielsen2002quantum}, which would imply that growing these types of discrete function approximations requires a vanishingly footnotesize amount of entanglement.

\begin{figure*}[t!]
  \centering
  \begin{tabular}[b]{c}
    \includegraphics[width=.27\linewidth,trim={0 0 1cm 0},clip]{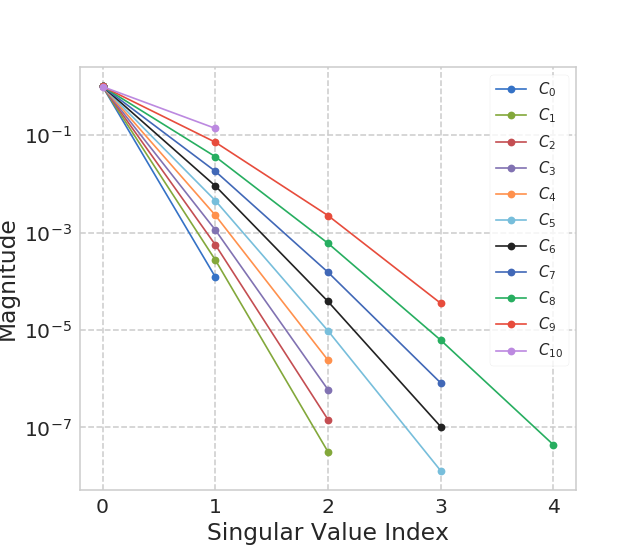} 
        \label{fig:gauss_1}\\
    \footnotesize (a) Gaussian $\mu=0$, $\sigma=1$
  \end{tabular} \qquad
  \begin{tabular}[b]{c}
    \includegraphics[width=.27\linewidth,trim={0 0 1cm 0},clip]{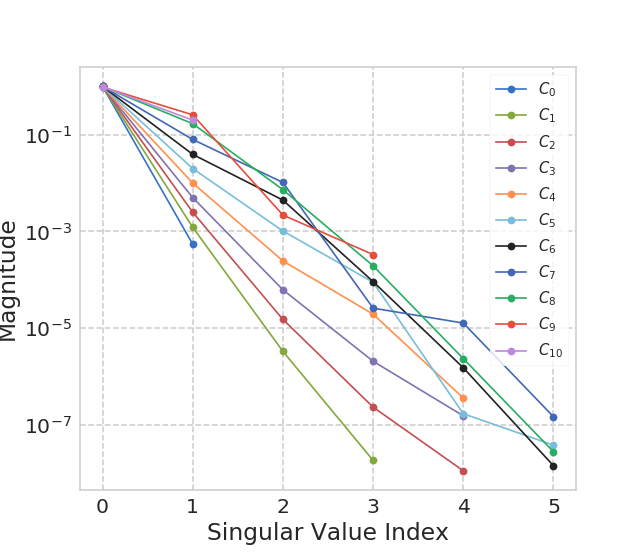} 
        \label{fig:lognormal_1}\\
    \footnotesize (b) Lognormal $\mu=1$, $\sigma=1$
   \end{tabular} \qquad
  \begin{tabular}[b]{c}
    \includegraphics[width=.27\linewidth,trim={0 0 1cm 0},clip]{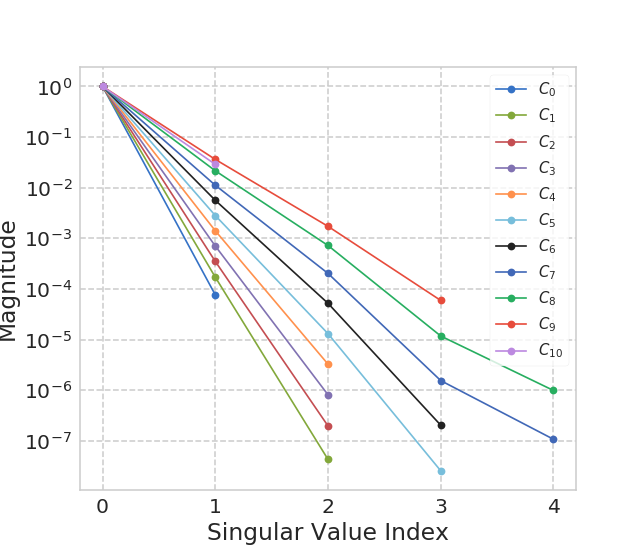} 
        \label{fig:lorentz_1}\\
    \footnotesize (c) Lorentzian $\mu=0$, $\sigma=1$
  \end{tabular} \qquad
    \begin{tabular}[b]{c}
    \includegraphics[width=.27\linewidth,trim={0 0 1cm 0},clip]{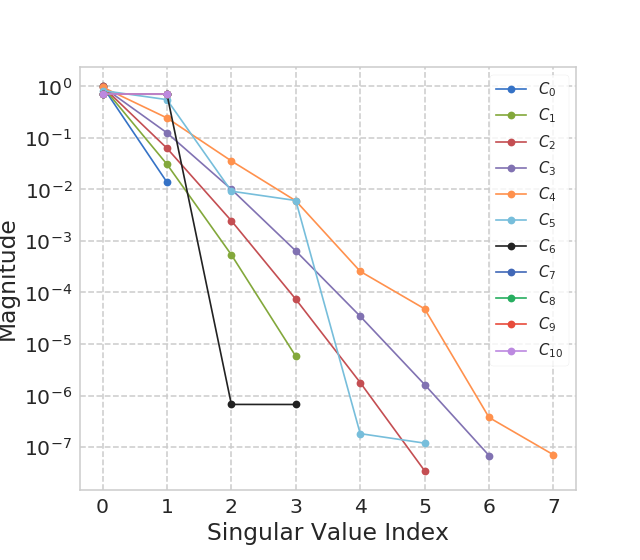}
        \label{fig:gauss_0.05}\\
    \footnotesize (d) Gaussian $\mu=0$, $\sigma=0.05$
   \end{tabular} \qquad
     \begin{tabular}[b]{c}
    \includegraphics[width=.27\linewidth,trim={0 0 1cm 0},clip]{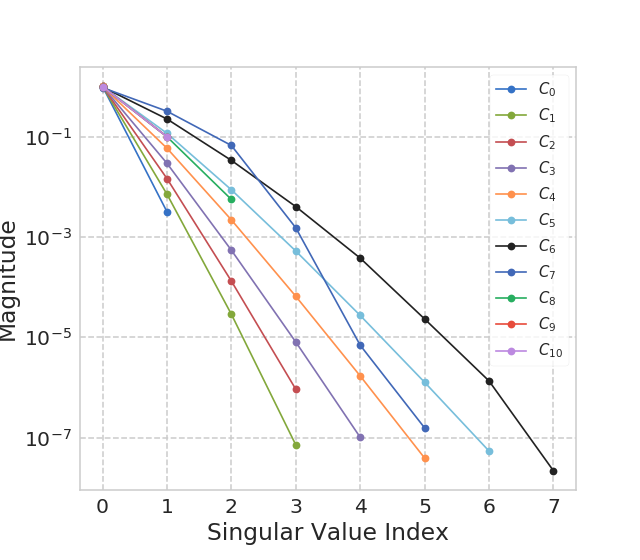} 
        \label{fig:lognormal_0.05}\\
    \footnotesize (e) Lognormal $\mu=1$, $\sigma=0.05$
   \end{tabular} \qquad
     \begin{tabular}[b]{c}
    \includegraphics[width=.27\linewidth,trim={0 0 1cm 0},clip]{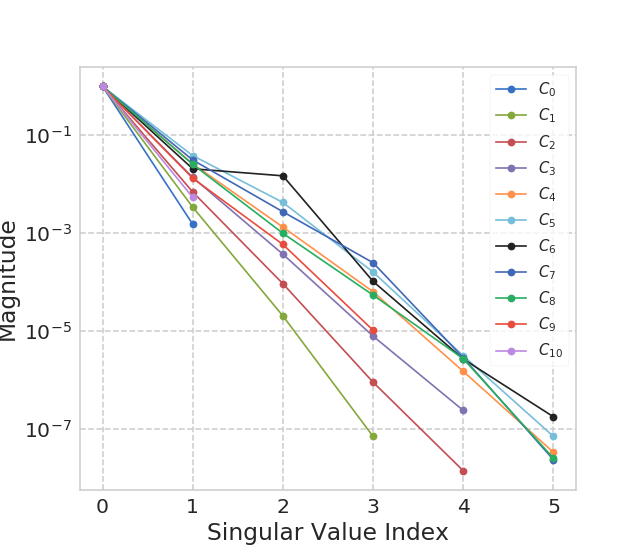} 
        \label{fig:lorentz_0.05}\\
    \footnotesize (f) Lorentzian $\mu=0$, $\sigma=0.05$
   \end{tabular} \qquad
  \caption{Spectral analysis of different SDR functions, discretized into a system of $N = 12$ qubits (color online)}
  \label{fig:sdr_spectra}
    \vspace{-0.5cm}
\end{figure*}

\subsection{Accuracy of low-$\chi$ Approximate MPS}
SDR functions can be connected with MPS representations through the concept of a low-rank matrix approximation. The maximum bond dimension $\chi$ of an MPS increases with the entanglement of a system \cite{schollwock2011density}, and because SDR functions have a maximum VNE increase bounded inverse-exponentially by system size of the discrete approximation, these functions are accurately approximated by low-$\chi$ MPS forms.
\begin{algorithm}[H]
	\caption{MPS-SVD}
	\label{alg:mps_svd}
	\algorithmicrequire{\;Target tensor $A$, System size $N$, Truncation parameter $\delta$ optional}\\
	\algorithmicensure{\;Approximate MPS $\boldvec{M}$ comprised of MPS cores $M^{[1]},\cdots,M^{[N]}$}
	\begin{algorithmic}[1]
	    \State $C \leftarrow A, \alpha_0 \leftarrow 1$
	    \For{ $k=1 \text{ to } N-1$ do}
	        \State $C \leftarrow \text{reshape}(C,[\alpha_{k-1}d_k,:])$
	        \State $U\Sigma V + E_{\delta} \leftarrow \delta$-truncated SVD($C$)
	        \label{alg:mps_svd:step:4}
	        \State $M^{[k]} \leftarrow \text{ reshape}(U,[\alpha_{k-1},d_k,\alpha_k])$
	        \State $C \leftarrow \Sigma V^T$
	   \EndFor
	   \State $M^{[N]} \leftarrow C$\\
	   \algorithmicreturn{ $\boldvec{M} = \{M^{[1]},\cdots,M^{[N]}\}$}  
	\end{algorithmic}
\end{algorithm}
The canonical algorithm for constructing an MPS representation of a target tensor is shown in Algorithm \ref{alg:mps_svd}, and was designed originally in \cite{oseledets2011tensor}. At a high level, the algorithm sweeps through the individual elements $1, 2, \cdots, N$ of the MPS, and at each site performs a singular value decomposition (SVD) of the target tensor. The resulting singular vectors are reshaped, labeled as the MPS core at this particular site.

Were we to perform a full SVD instead of $\delta$-truncation in step \ref{alg:mps_svd:step:4} of Algorithm \ref{alg:mps_svd}, the constructed MPS would be exact. If however we approximate the \textit{unfolding matrix} $C_k$ at step $k$, and leave out singular values that fall beneath some threshold $\delta$, we incur an error $\varepsilon \leq \sqrt{\sum_{j=1}^N \varepsilon^2_j}$ where each $\varepsilon_j$ is the individual truncated-SVD error given by $||E_{\delta}||_F$ in the algorithm, see \cite{oseledets2011tensor} \S 2.2. 

The optimality of the resulting MPS is given by the Eckart-Young theorem \cite{eckart1936approximation,golub2013matrix}, from which we know that the best rank $k$ approximation to a matrix $A \in \mathbb{R}^{m \times n}$ is given by considering the first $k$ singular values and vectors, and the error under the Frobenius norm is equal to the total of omitted singular values: 
\begin{align*}
    A_k &= \sum_{i=1}^k \sigma_i u_i v_i^T \quad \rightarrow \quad 
    ||A - A_k||_F = \sqrt{\sum_{j=k+1}^N \sigma^2_{j}}
\end{align*}
This implies that the MPS constructed in $\delta$ truncated Algorithm \ref{alg:mps_svd} is an optimal MPS for the specified $\delta$; the MPS core formed by approximating unfolding matrix $C_k$ in step \ref{alg:mps_svd:step:4} is optimal.

With these conditions, we can estimate the accuracy of bounded-$\chi$ approximate MPS representations of SDR functions. Connecting the $\delta$-truncated SVD error with the error from Algorithm \ref{alg:mps_svd}, we see that the approximate MPS $\boldvec{M}$ with bond dimension bounded at $\chi$ has Frobenius error upper bounded by the sum of all squared omitted singular values of the unfolding matrices:
\begin{align}
    ||A - \boldvec{M}||^2_F &\leq \sum_{j=1}^N \varepsilon_j^2
    = \sum_{j=1}^N \bigg(\sum_{k=\chi + 1}^{\text{dim}(C_j)}\sigma_{k}^2(C_j)\bigg)
\end{align}

Based on this, we conjecture and show empirically that the spectra of unfolding matrices decays exponentially for discretized SDR functions, potentially because the entropy grows inverse-exponentially as qubits are added.  This implies the existence of accurate low $\chi$ MPS approximations, and we show high-accuracy approximations even for $\chi=2$. Unfolding matrix spectra are shown in Figure \ref{fig:sdr_spectra}, while numerical evidence supporting the exponential decay of equation (\ref{eq:sv_exp_decay}) is shown in Figure \ref{fig:exp_decay_rates}. 

We can estimate the accuracy of $\chi=2$ MPS approximations to SDR functions by modeling the spectra of unfolding matrices with exponential decay. Allowing for each $C_j$ unfolding matrix to follow a distinct exponential decay, we can formulate an exponential univariate multiple regression with the model shown in equation (\ref{eq:sv_exp_decay}).
\begin{align}
     \sigma_k^j &= \alpha_j e^{-\beta_j k}
    \label{eq:sv_exp_decay}
\end{align}

We have a two-parameter univariate exponential decay model for the spectrum of each $C_j$, where the $j$-th unfolding matrix spectrum is characterized by empirically fit parameters $\alpha_j, \beta_j$. Under this model, we can calculate the normalized upper bound of the error of an MPS approximation with bounded bond-dimension $\chi$, shown in equation (\ref{eq:exp_decay_bounds}), where we have assumed for simplicity of analytics that all $\alpha, \beta$ terms are approximately equal. This allows us to estimate that for $\chi=2$ and all $\beta \geq 1.152$, there will exist an MPS representation with greater than $99\%$ normalized accuracy. 
\begin{align}
    \frac{||A-\boldvec{M}||_F^2}{||A||_F^2} &\leq \frac{\sum_{j=1}^N \bigg(\sum_{k=\chi + 1}^{\text{dim}(C_j)}(\alpha_je^{-\beta_jk})^2\bigg)}{\sum_{j'=1}^N \bigg(\sum_{k'=1}^{\text{dim}(C_{j'})}(\alpha_{j'}e^{-\beta_{j'}k})^2\bigg)}\nonumber \\
    &= \frac{e^{\beta(N-2)}(e^{2\beta}-e^{-2\beta})}{e^{\beta N}-e^{-\beta N}}\nonumber\\
    &= e^{\beta(N-\chi)}\text{csch}(\beta N)\sinh{(\chi\beta)}
    \label{eq:exp_decay_bounds}
\end{align}
\subsection{Numerical SDR Spectral Analysis}
\begin{figure*}[t!]
\centering
  \begin{tabular}[b]{c}
    \includegraphics[width=.4\linewidth,trim={-0cm 0 0cm 0cm},clip]{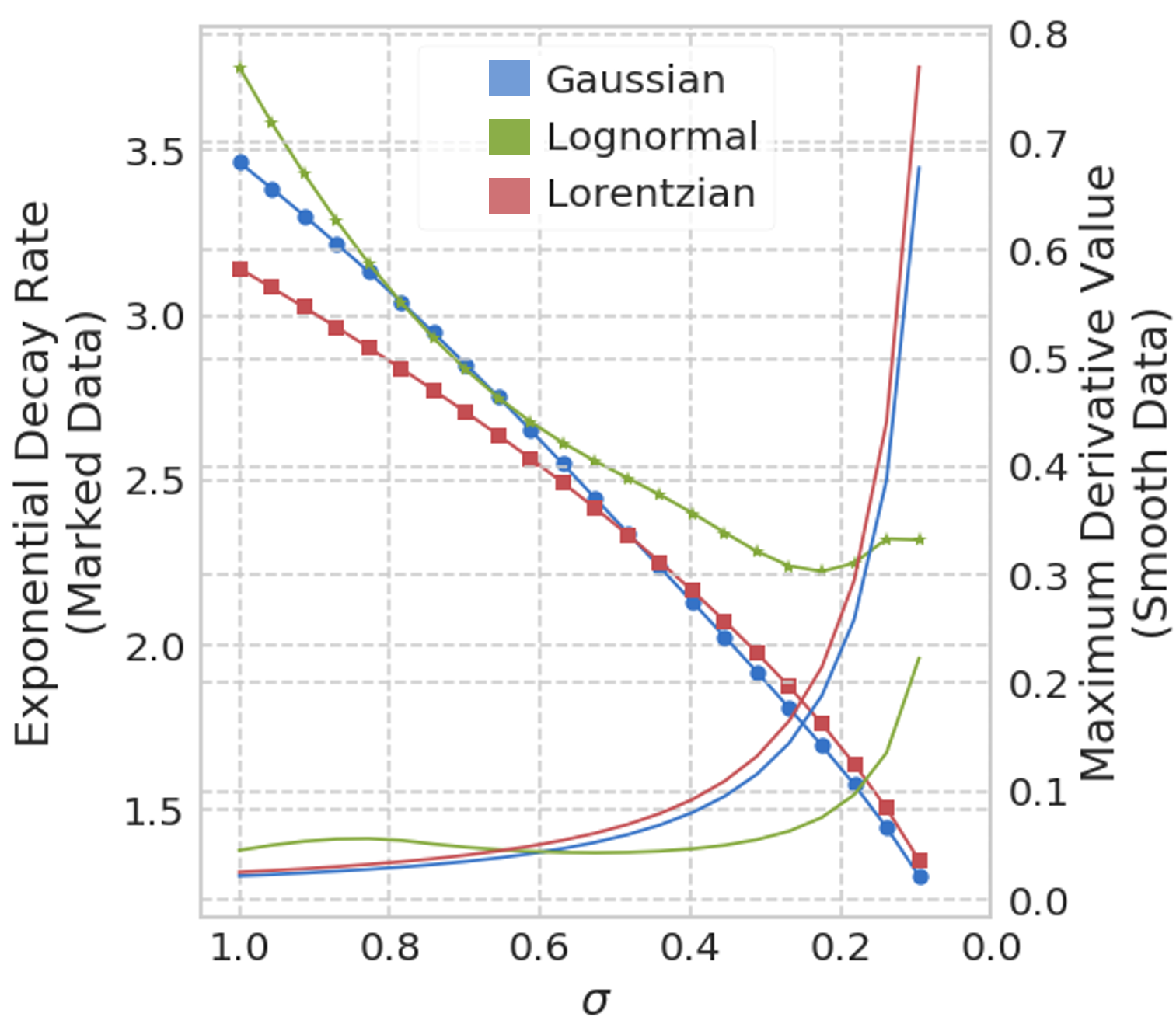} 
        \label{fig:exp_decay_rates_sigma}\\
    \footnotesize (a) 
  \end{tabular} \qquad
  \begin{tabular}[b]{c} 
    \includegraphics[width=.4\linewidth,trim={0 0 0 0},clip]{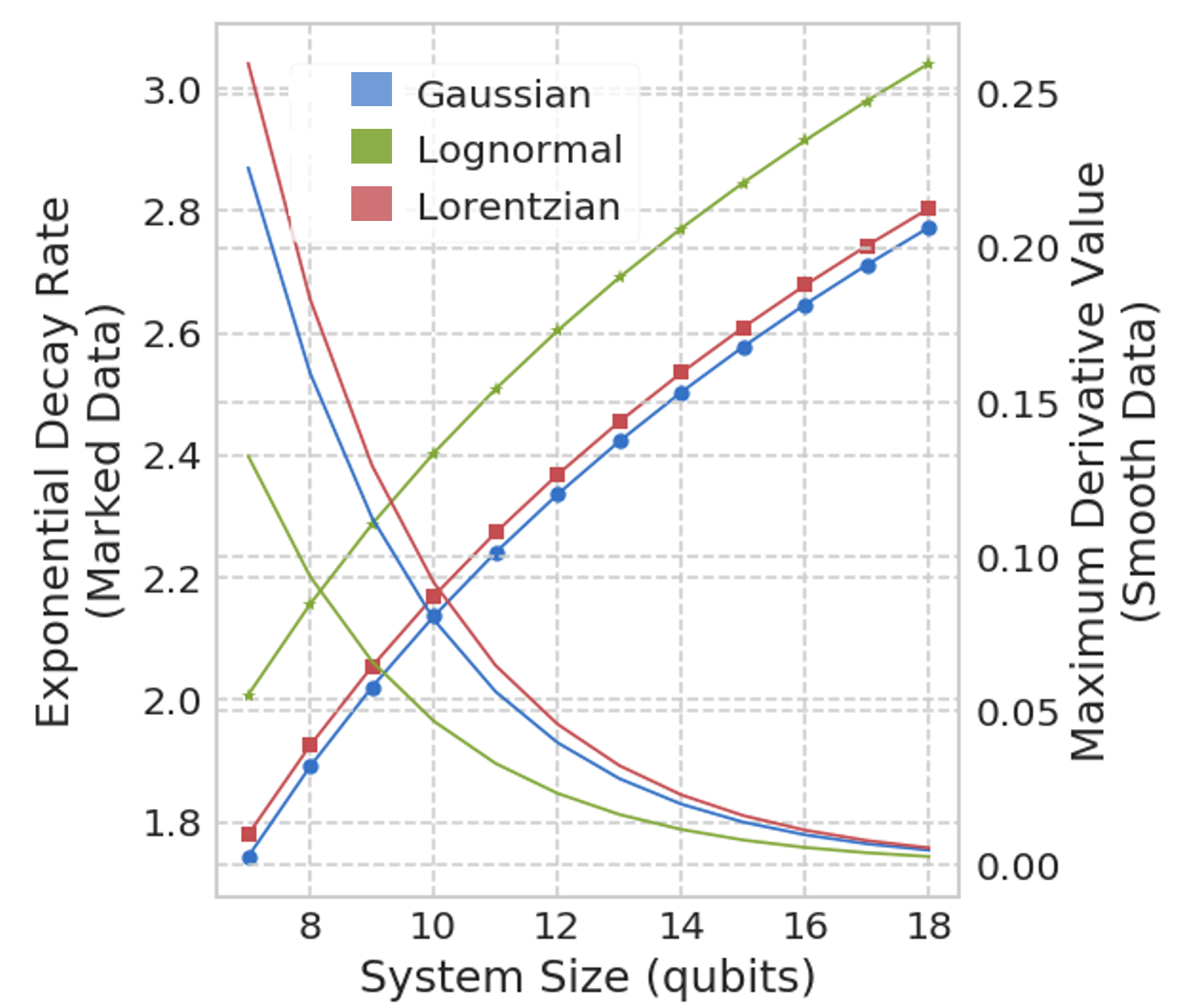}
        \label{fig:exp_decay_rates_system}\\
    \footnotesize (b)
   \end{tabular} \qquad
  \caption{Marked data indicates empirically fit exponential decay rates across all unfolding matrix spectra for each SDR distribution, plotted against (a) decreasing standard deviation -- or increased "squeezing" and (b) increasing system size for fixed $\sigma=0.4$. Smooth lined data indicates the maximum derivative achieved for each of the distributions. For all distributions, as the function is squeezed the maximum derivative increases while the exponential decay rate of the singular values decreases. The lognormal distribution qualitatively changes shape around $\sigma \leq 0.2$, leading to a change in the maximum derivative resulting in a phase change of the decay rate of the singular values. We see in (b) an increase in the rate of exponential decay as systems increase in size, indicating that $\chi=2$ MPS approximators remain effective as systems are increased in size.} 
  \label{fig:exp_decay_rates}
    \vspace{-0.5cm}
\end{figure*}
The validity of a singular value decay rate following the model of equation (\ref{eq:sv_exp_decay}) can be numerically estimated. Empirically, we find that the spectra are well modeled by this form, and estimated $\beta$ values are often larger than this threshold. We highlight univariate Gaussian, lognormal, and Lorentzian distributions, as they are representative distributions commonly used in applications, and each is discretized across a bounded support interval. The probability density functions of these distributions are shown in equations (\ref{eq:gaussian}), (\ref{eq:lognormal}), and (\ref{eq:lorentzian}).

Figure \ref{fig:sdr_spectra} shows the unfolding matrix spectra for Gaussian, lognormal, and Lorentzian distributions with varying degrees of squeezing. \textit{Squeeze} here refers to the inverse of the standard deviation of the distribution: $\sim 1/\sigma$. The term refers to how tightly centered the distributions are around the means. Loosely, as the standard deviations decrease for these functions, the maximum derivative of the functions obtained on the supported domains increases, in that tightly squeezed states with low $\sigma$ have higher maximum derivatives than their high-$\sigma$ counterparts. This likely contributes to the fact that low-rank MPS approximations have difficulty capturing these highly localized features.
\begin{align}
    p_G(x;\mu,\sigma) &= \frac{1}{\sqrt{2\pi}\sigma}e^{(-(x-\mu)^2)/2\sigma^2}\label{eq:gaussian}\\
    p_{Ln}(x;\mu,\sigma) &= \frac{1}{x}p_G(\log(x))\label{eq:lognormal}\\
    p_L(x;\mu,\sigma) &= \frac{\sigma}{2\pi}\frac{1}{(x-\mu)^2+\sigma^2}\label{eq:lorentzian}
\end{align}

The joint exponential decay rate is found by fitting the model of equation (\ref{eq:sv_exp_decay}) to all of the composite spectra. We find that for these distributions, the decay rates are above $1.152$ for all but the most "squeezed" Gaussian distributions. Results are shown in Figure \ref{fig:exp_decay_rates}. As these distributions become tighter with footnotesizeer standard deviation, these functions gain larger and larger derivatives through the supported domain, which likely prevents low-rank approximations from capturing these highly local features completely. Their unfolding matrix spectra decay slower and slower as well. Empirically, good $\chi=2$ MPS representations of these distributions can be formed with greater than $99\%$ accuracy as is predicted by our analytical model, so long as the standard deviations are moderately valued, holding the discretization domain constant. We also see that as systems increase in size, the value of the maximum derivative decreases, and the exponential decay rates actually increase. This indicates that $\chi=2$ MPS likely remain good if not better approximations for states discretized over large systems.

\section{Techniques}
\label{sec:techniques}

The core of our algorithm rests on four main techniques: piecewise polynomial function approximation, MPS arithmetic, iterative MPS compression, and quantum gate extraction from MPS representations. 
\subsection{Piecewise polynomial function approximation MPS}
\label{sec:techniques_pp_approx}
In many cases, matrix product states are used to encode low-rank approximations to data which do not have a known analytical form. In these cases MPS forms can be constructed using exact construction as in Algorithm \ref{alg:mps_svd}. They can also be approximately constructed using algorithms that subsample a portion of the domain and interpolate \cite{oseledets2010ttcross}, extract dominant singular values exactly \cite{lee2014very,dolgov2014computation}, or estimate the dominant singular values potentially with randomized algorithms \cite{huber2017randomized,batselier2018computing,che2019randomized}. Recent work \cite{dolgov2020approximation} applies these techniques to develop a method in this fashion for sampling potentially exotic multivariate probability distributions.  

In our case, we are presented with an analytical form of the state we are constructing. Many functions with analytic forms can be exactly written down in a matrix product state, as shown in \cite{oseledets2013constructive}. However, a technique to do so requires that these functions are \textit{rank}-$r$ \textit{separable}. This means that these functions can be written as equation (\ref{eq:separable}), for some fixed $r$ value. Unfortunately, this property does not hold for many probability density functions. It does hold however for degree-$p$ polynomials, as in equation (\ref{eq:polynomial})
\begin{align}
    f(x+y) &= \sum_{\alpha=1}^r u_{\alpha}(x)v_{\alpha}(y)
    \label{eq:separable}\\
    f(x) &= \sum_k^p a_kx^p
    \label{eq:polynomial}\\
\end{align}
An explicit form of discretized functions of the form (\ref{eq:polynomial}) can be written as:
\begin{align}
    \label{eq:piecewise_poly_mps}
    f\bigg(\sum_{k} t_k\bigg) &= M_{\alpha_1,\alpha_2}^{[1],s_0}M_{\alpha_2,\alpha_3}^{[2],s_1}...M_{\alpha_{N},\alpha_{N+1}}^{[N],s_{N}}\\
    \phi_s(x) &= \sum_{k=s}^p a_k \binom{k}{s} x^{k-s}\nonumber\\
    M^{[k],t_k}_{i,j} &= \begin{cases}\binom{i}{i-j}x^{i-j}&i\geq j\\0 & \text{otherwise} \end{cases}\nonumber\\
    \vspace{-0.5cm}
\end{align}
where we have for the first and last tensors:
\begin{align*}
    M^{[1],t_1} &= \bigg(\phi_0(t_1), \phi_1(t_1), \cdots, \phi_p(t_1)\bigg)\nonumber\\
    M^{[N],t_N} &= \bigg(1, t_N^1, t_N^2, \cdots, t_N^p\bigg)\nonumber
\end{align*}
These MPS forms have bounded $\chi \leq p+1$, see \cite{oseledets2013constructive}, \S 6.

Motivated by this, we derive novel MPS forms of \textit{piecewise} polynomial (PP) functions with bounded support on a subregion of the gridded domain. Specifically, for a domain $D = [a,b]$ and subdomain $a', b'$ such that $a < a' < b' < b$, a polynomial function with support on domain $D' = [a',b']$ can be written as:
\begin{align}
    f(x) = \begin{cases}\sum_k^p a_kx^k & a' \leq x \leq b'\\0 & \text{otherwise}\end{cases}
    \label{eq:piecewise_polynomial}
\end{align}

Based on a binary encoding of the original domain, subdivide the domain into a set defined by \textit{support-bit} $k$ ordered from the left. This creates $2^k$ different regions, each defined as:
\begin{align}
    D_j= \bigg[a+j\frac{2^kL}{h},a+j\frac{2^{k}L}{h}+\frac{2^{N-j}L}{h}\bigg)
\end{align}
Here we use the encoding provided in equation (\ref{eq:index_domain_map}), where the $j^{\text{th}}$ polynomial is supported on the region indexed by $j$, and assert that the last region is inclusively bounded. This creates a uniform gridding of the domain into $2^k$ evenly spaced partitions, each of which then supports a single function approximating polynomial.

The $j^{\text{th}}$ polynomial in a piecewise approximation over a support-bit $k$ divided domain can be referred to as $g_j(x)$, and can be written into an MPS as in form (\ref{eq:piecewise_poly_mps}), with explicit zeroing out of the tensors that correspond to domain values outside support. To do this, we write out $\bm{b_j} = \{b_1, b_2, \cdots, b_k\}$ a binary representation of the index $j$ using exactly $k$ bits. Then, for each tensor $i$ from $1 \cdots N$, we zero out the component of the tensor that is unsupported in the binary encoding of the domain. Explicitly, for all $1 \leq i \leq k$: 
\begin{align}
    \label{eq:k_zeroed}
    M^{[i],t_i} : 
    \begin{cases}
    M^{[i],t_i=1} = \bm{0^{p+1 \times p+1}} & \text{if } b_i = 0\\
    M^{[i],t_i=0} = \bm{0^{p+1 \times p+1}} & \text{if } b_i = 1
    \end{cases}
\end{align}
where the remaining $M^{[i],t_i}$ for $i > k$ are all unchanged. Equation (\ref{eq:k_zeroed}) enforces that the polynomial $f_j(x)$ is zeroed out for any domain value that lies outside of the range $D_j$.

With this explicit form of a bounded-support polynomial, we can write the total MPS of a piecewise polynomial function used as a function approximator. A piecewise polynomial approximation function $g(x)$ with support on domain $D$ is constructed by subdividing $D = \{D_1, D_2, \cdots, D_{2^k}\}$ into $2^k$ subregions, and fitting $2^k$ possibly-distinct piecewise polynomials to the function $g(x)$, with each polynomial supported on a single subregion. Together, this forms a piecewise polynomial approximation $\Tilde{g}(x)$ to $g(x)$:
\begin{align}
    \Tilde{g}(x) = \begin{cases}
                    g_1(x) & x \in D_1\\
                    \cdots \\
                    g_k(x) & x \in D_k\\
                    \end{cases}
\end{align}
Here we do not require the functions be continuous at the endpoints of subregions.

With a function approximation written down this way, we can iteratively construct a series of $2^k$ MPS forms corresponding to each of the approximating polynomials.

\begin{figure}[t!]
    \centering
    \includegraphics[width=0.8\linewidth]{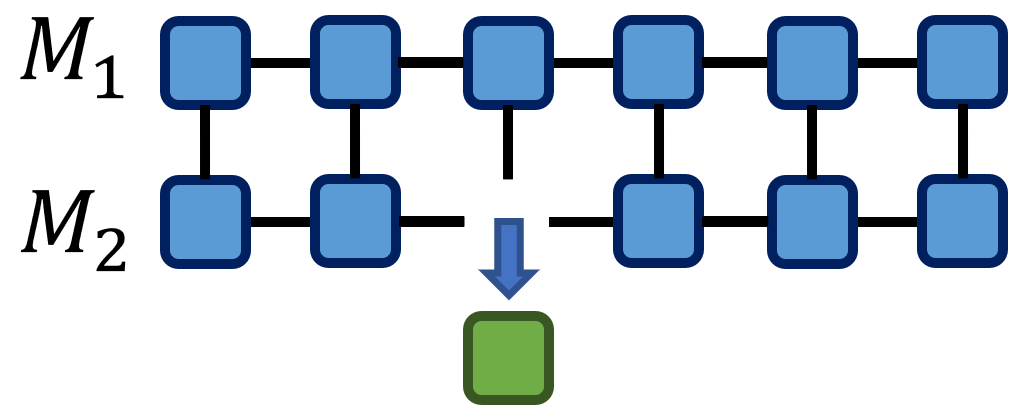}
    \caption{Evaluating the partial derivative of the overlap between two matrix product states, at site $2$, as zero-indexed from the left. Full contraction of $\bm{M_1}$ and $\bm{M_2}$ is completed, omitting site $2$. The resulting tensor system is solved for the optimal site-$2$ tensor that maximizes value of the overlap, with normalization constraints. This new tensor replaces the original site-$2$ tensor.}
    \label{fig:gradient_overlap}
    \vspace{-0.5cm}
\end{figure}

\subsection{Matrix product state arithmetic}
\label{sec:techniques_mps_addition}
Once we form $2^k$ MPS forms, we can combine them using the arithmetic properties of matrix product states. Specifically, two matrix product states can be added together as:
\begin{align*}
    \bm{M_1} + \bm{M_2} &= \bm{M_3} = M_3^{[1],s_1}M_3^{[2],s_2}...M_3^{[N],s_{N}}
\end{align*}
where each $M_3^{[i]}$ term is a block diagonalization of the corresponding terms in each summand:
\begin{align*}
M_3^{[i],s_i} = \text{diag}\bigg(M_1^{[i],s_i}, M_2^{[i],s_i}\bigg)
\end{align*}
appropriately adjusting for single-dimensional row and column vector endpoint cores. Upon contraction, the result is the addition of the two encoded functions:
\begin{align*}
    \bm{M_3}(x) &= \sum_{\bm{\alpha_1}} \prod_{i=1}^{N}M_{1,\alpha_{i},\alpha_{i+1}}^{[i],s_i}+\sum_{\bm{\alpha_1}} \prod_{i=1}^{N}M_{2,\alpha_{i},\alpha_{i+1}}^{[i],s_i}\\
    &=\bm{M_1}(x)+\bm{M_2}(x)
\end{align*}
Using this property, we can combine the $2^k$ MPS forms defined in a piecewise polynomial function approximation. Each of the constituent MPS forms have a maximum bond-dimension defined by the degree of the encoded polynomial: $\chi_j \leq p_j+1$. MPS addition in this way grows the rank of the resulting MPS by exactly the ranks of the constituent MPS. Because of this, the MPS formed by the addition of $2^k$ degree-$p$ piecewise polynomial MPS forms has rank $\chi_{\text{total}} = 2^k(p+1)$.

\subsection{Iterative MPS Compression}

\label{sec:techniques_mps_compress}
Large-$\chi$ matrix product states can be compressed into a lower-$\chi$ MPS by an iterative method, focusing on a single core at a time. Following \cite{schollwock2011density}, the optimal approach to compress an MPS $\bm{M_1}$ of rank $\chi_1$ into $\bm{M_2}$ of rank $\chi_2 < \chi_1$ is to begin with an ansatz for $\bm{M_2}$ of rank $\chi_2$, and change each core $M_2^{[k],s_k}$ iteratively. For core $k$, the update rule follows from the gradient of the overlap between both states, calculated with respect to core $k$. This gradient is of the form:
\begin{align}
    \frac{\partial\braket{\bm{M_1}|\bm{M_2}}}{\partial M_2^{[k]}} = \sum_{\bm{\alpha},\bm{S}}\bigg[\bigg(\prod_{i \in \mathcal{I}}M_1^{[i],s_i}\bigg)^{\dagger}\prod_{j \in \mathcal{I}/\{k\}}M_2^{[j],s_j}\bigg]
    \label{eq:partial_derivative_overlap}
\end{align}
which corresponds to a full pairwise contraction of the conjugate of each core in $\bm{M_1}$ with the corresponding core of $\bm{M_2}$, \textit{omitting} the $k$-th core in $\bm{M_2}$. As such, above the index sets $\mathcal{I}$ are defined as the ordered set $\{N, N-1, \cdots, 1\}$. In graphical notation this is simplified as shown in Figure \ref{fig:gradient_overlap}.

The iterative compression algorithm evaluates equation (\ref{eq:partial_derivative_overlap}) at each site $k$, and calculates the optimal core $k$ to replace the existing $k$-th core. This calculation corresponds to solving a $\big(\chi_2^2 \times \chi_2^2\big)$ dimensional linear system, and using factorizations presented in \cite{schollwock2011density} can be performed in $\mathcal{O}(\chi_2^3)$ time. In practice then, this algorithm can be computationally bounded at $\mathcal{O}(N\chi_2^3)$ time, where a fixed number of sweeps and solutions are performed over all $N$ cores.

\begin{algorithm}[H]
	\caption{SDR Function Encoding}
	\label{alg:main_algorithm}
	\algorithmicrequire{\;Target SDR function $f(x)$, System size $N$, Support-bit $k$, Domain $D = [a,b]$, Accuracy parameter $\varepsilon$ optional}\\
	\algorithmicensure{\;Quantum circuit $U = g_1g_2\cdots g_k$ such that \\$U\ket{0} = \ket{\psi_{f(x)}}$}
	\begin{algorithmic}[1]
	    \State $R \leftarrow \{D_1, D_2, \cdots, D_{2^k}\}$ \Comment{binary subdivided domain}
	    \State $\Tilde{g}(x) \leftarrow \{g_j(x)\ |\ \forall j \in [1,2^k]\}$ \Comment{PP approx of $f(x)$}
	    \For{ $j=1 \text{ to } 2^k$ do}
	        \State $M_j \leftarrow \text{MPS}(g_j(x))$ \Comment{ MPS encoding of $g_j$ \S \ref{sec:techniques_pp_approx}}
	   \EndFor
	   \State $M_T \leftarrow \sum_k M_k$ \Comment{MPS summation \S \ref{sec:techniques_mps_addition}}
	   \State $\Tilde{M}_T \leftarrow \text{IterCompress}(M_t)$ \Comment{MPS compression \S \ref{sec:techniques_mps_compress}}
	   \State $G \leftarrow \text{GateExtraction}(\Tilde{M}_T)$\Comment{MPS to q-gates \S \ref{sec:techniques_gate_extraction}}\\
	   \algorithmicreturn{ $G = g_1, g_2, \cdots, g_k$}  
	\end{algorithmic}
\end{algorithm}

\subsection{Accurate linear-depth circuits}
\label{sec:techniques_gate_extraction}
Once a suitable matrix product state has been constructed, a technique developed recently in \cite{ran2020encoding} can be used to directly convert the MPS into a set of one and two-qubit gates. This is performed by calculating the ``matrix product disentangler" $\hat{U}$ with the property that it acts on the state $\ket{\psi}$ encoded by the MPS and creates the vacuum state.

The procedure for construction of this unitary operator acts on each MPS core $1 < k < N$ and forms two-qubit gate $G^{[k]}$:
\begin{align*}
    G_{0js_kl}^{[k]} = M^{[k],s_k}_{j,l}
\end{align*}
This forms half of the required elements for the two qubit operator $G^{[k]}$, and the other half are chosen by selecting $(d^2-d)$ orthonormal vectors in the kernel of $M^{[k]}$. This fills in the two qubit operator, and results in a unitary gate. Sites $1$ and $N$ are filled in similarly, adjusting for specific dimensional constraints. The result is a set of $N+1$ unitary quantum operations, $N$ of which are two qubit gates, that form a serialized circuit of depth linear in system size: $N+1$. Details of the procedure are shown clearly in \cite{ran2020encoding}. These circuits can be decomposed into canonical gates using $7N+1$ single qubit gates and $3N$ two qubit gates, at a depth of $\sim 6N$, utilizing standard decompositions \cite{coles2018quantum}.

\begin{figure}[t!]
  \centering
  \begin{tabular}[b]{c}
    \includegraphics[width=0.8\linewidth,trim={0cm 0 0cm 0},clip]{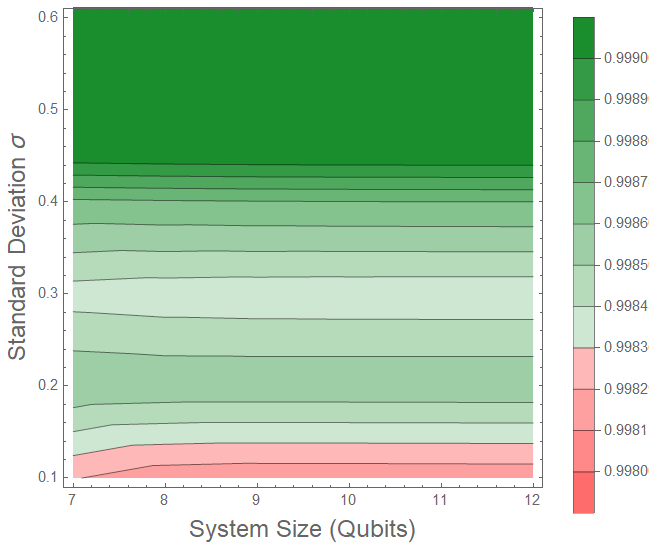} 
        \label{fig:circuit_fidelity_all}
   \end{tabular} \qquad
  \caption{Scaling of circuit fidelities with the standard deviation of the target probability distribution, across low and intermediate $\sigma$ for Gaussian, lognormal, and Lorentzian distributions. Circuits are constructed using support-bit $k=3$, dividing the domain into $2^k=8$ pieces. All circuits construct states with greater than $99\%$ fidelity through $\sigma \geq 0.1$, and exceed $99.9\%$ for all system sizes and all distributions above $\sigma=0.44$. Below, circuit fidelities decay with the ability to approximate the distributions with low rank $\chi=2$ MPS forms, in agreement with predictions from equation (\ref{eq:exp_decay_bounds}). In each distribution, we set $\mu = 1$ and bound support on domain $D = [0,2]$, with $D=[\varepsilon,5]$ for the lognormal distribution to capture relevant features. Within the $\sigma=0.3$ region, there are changes in the Gaussian and lognormal distributions that make approximation by a $\chi=2$ MPS more difficult with $8$ piecewise polynomials.}
  \label{fig:results_analysis}
    \vspace{-0.5cm}
\end{figure}

\section{Algorithm}
\label{sec:algorithm}
All of the techniques from Section \ref{sec:techniques} are combined into Algorithm \ref{alg:main_algorithm}, which has a time complexity of $\mathcal{O}(N\chi^3) = \mathcal{O}(8N)$ for $\chi=2$ MPS approximations. Proving this requires analyzing each component of the algorithm.


Procedures in lines 1, 4, and 6 are constant time components. 
Each of the MPS encodings of line 4 can be performed in parallel, as they are each independent and are a constant time operations, following the explicit analytical form prescribed in equation (\ref{eq:piecewise_poly_mps}) and truncating with equation (\ref{eq:k_zeroed}). This is an important component of the algorithm, and the number of regions $2^k$ is a constant that often provides sufficient accuracy when chosen to be small (e.g. 8). MPS summation in line 6 is also constant-time as it is reorganizes the tensors into block diagonalizations of the constituent piecewise-supported MPS. 

Constructing the piecewise approximation of the function in line 2 has complexity that reflects the method used to do the approximating. Each subregion $D_j \in R$ is independent, so all $2^k$ approximations can be performed in parallel. A single approximation over region $j$ by a bounded degree-$p$ polynomial can be performed with complexity that scales with the number of distinctly sampled points on each subregion. For a gridding of domain $D_j$ into $L$ points, the least squares polynomial regression can be performed in $\mathcal{O}(p^2L)$, where both $p$ and $L$ are constants chosen and customizable to particular target functions. 

The iterative MPS compression of line 7 is the dominant contributing factor to the complexity of the algorithm, requiring $\mathcal{O}(N\chi^3)$ where we are targeting states with $\chi=2$, reducing this to $\mathcal{O}(8N)$. This complexity bound arises as the compression can be simplified into computing the optimal single-tensor that solves $N$ distinct overlaps between a $\chi_1=p+1$ state and the optimized $\chi_2=2$ state, each of which amounts to solving a $(\chi_2^3 \times \chi_2^3)$ linear system, after accounting for useful properties afforded by the MPS representation \cite{schollwock2011density}.

Gate extraction \cite{ran2020encoding} is also a linear time operation, requiring the inversion of $N$ $\chi_2 \times \chi_2$ matrices to complete each quantum gate. This can be done naively in $\mathcal{O}(\chi_2^3)$ time, and at best $\mathcal{O}(\chi_2^{2.373})$ \cite{burgisser2013algebraic}. Once again, all $N$ matrices can be inverted in parallel, reducing this complexity to $\mathcal{O}(N\chi^{2.373})$. 

Altogether, iterative MPS compression is the dominant computational cost of the algorithm when simple function approximation and optimized matrix inversion techniques are used, resulting in a time complexity for $\chi=2$ targets of $\mathcal{O}(8N)$.

\section{Analysis and Results}
\label{sec:results}

 \begin{figure}[t!]
    \centering
    \includegraphics[width=0.8\linewidth]{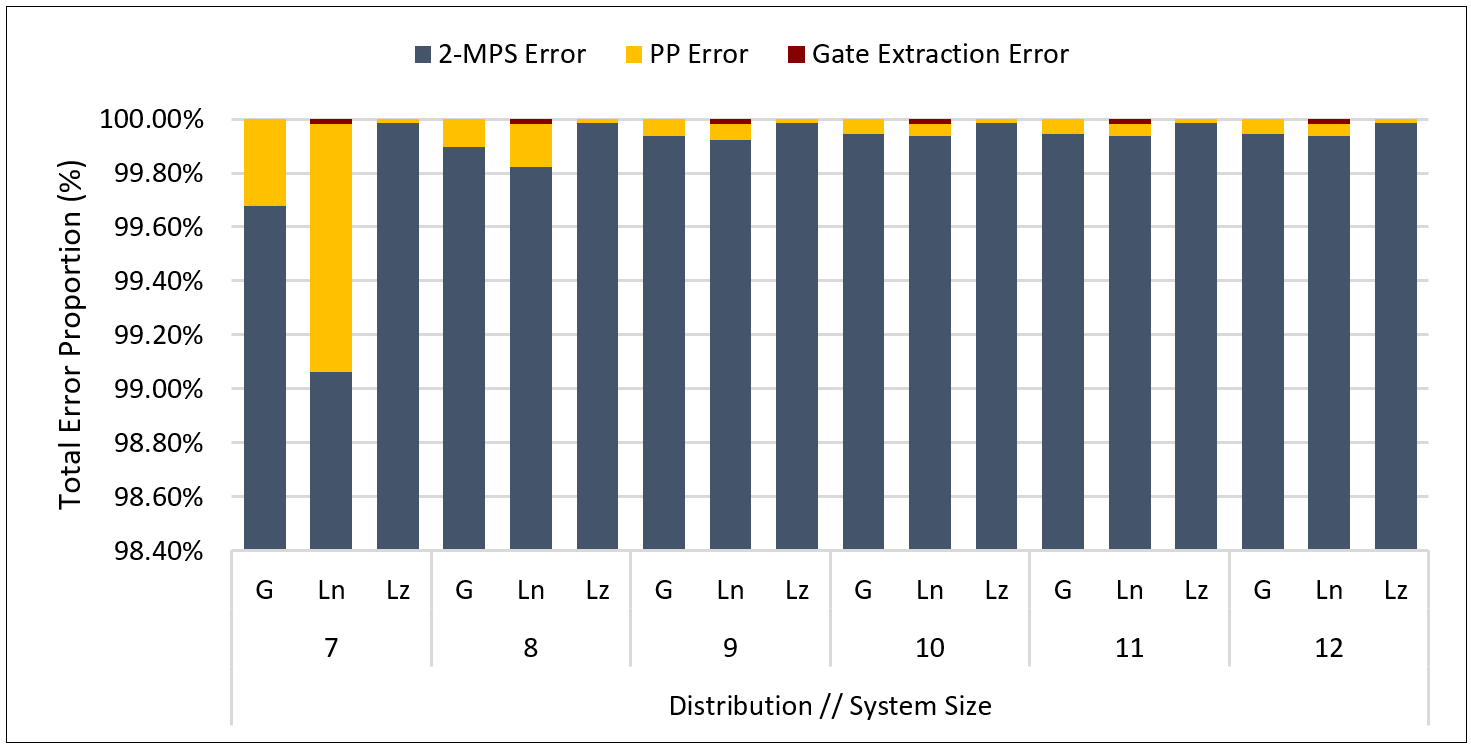}
    \caption{Decomposing the total error of each state construction by source. G, Ln, and Lz correspond to Gaussian, lognormal, and Lorentzian distributions, respectively. Configurations correspond exactly to those constructed in Figure \ref{fig:results_analysis}, for fixed $\sigma=0.1$.}
    \label{fig:stacked_error}
\end{figure}
\begin{figure}[t!]
    \centering
    \includegraphics[width=0.8\linewidth]{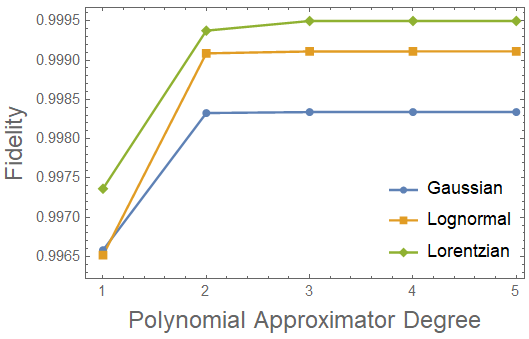}
    \caption{Accuracy of Algorithm \ref{alg:main_algorithm} sweeping through degrees of polynomial approximators, for a fixed sample size within subregions and for distributions with fixed $\sigma=0.3$. Diminishing returns are seen for polynomials higher than order $2$, which is a property of these specific functions.}
    \label{fig:degree_study}
    \vspace{-0.5cm}
\end{figure}

\subsection{Performance analysis and Numerical Results}

\begin{figure*}[t!]
  \centering
  \begin{tabular}[b]{c}
    \includegraphics[width=.27\linewidth,trim={-0cm 0 0cm 0cm},clip]{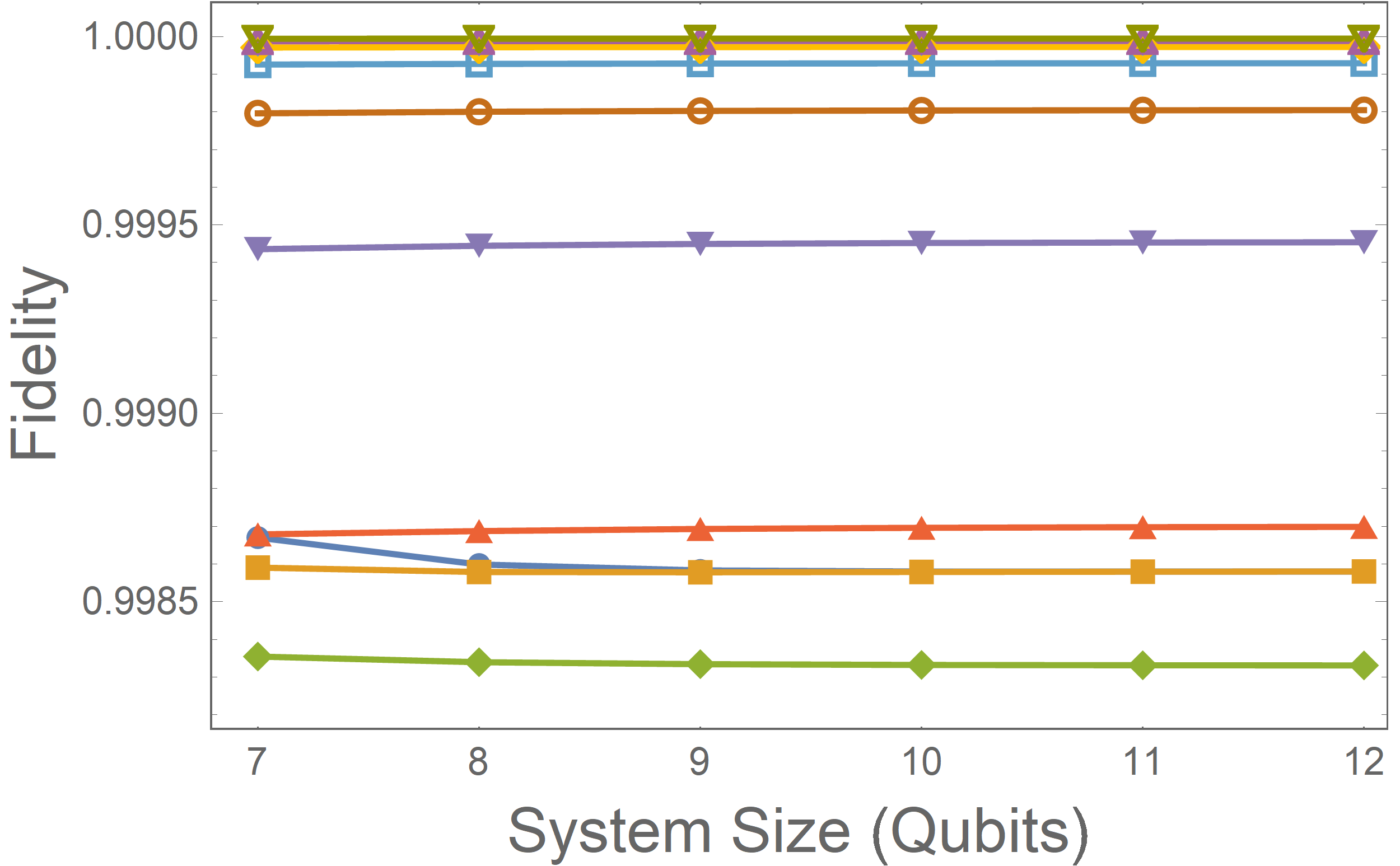} 
        \label{fig:gaussian_opt}\\
    \footnotesize (a) Gaussian SVD $\chi=2$ MPS
  \end{tabular} \qquad
  \begin{tabular}[b]{c} 
    \includegraphics[width=.27\linewidth,trim={0 0 0 0},clip]{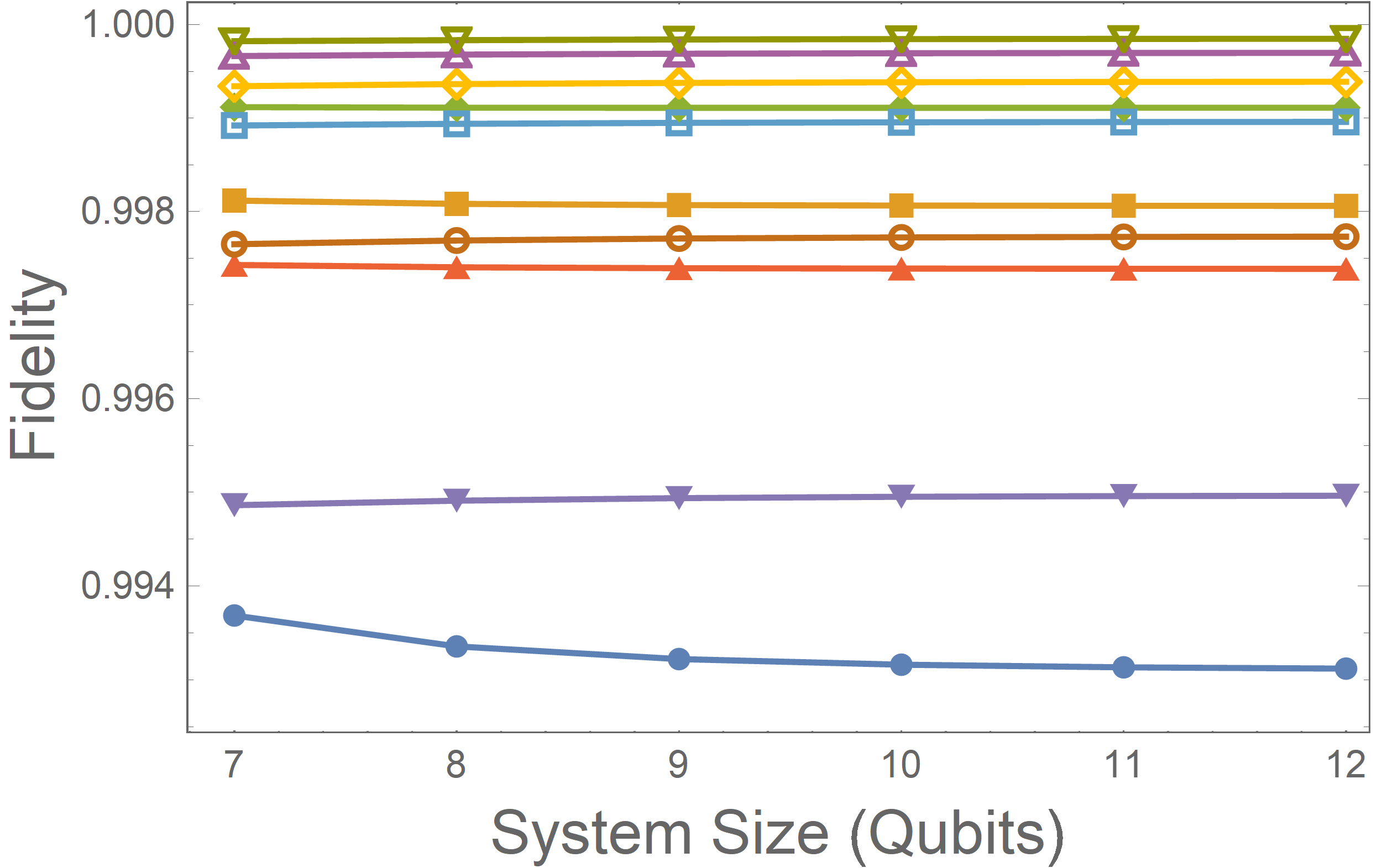}
        \label{fig:lognormal_opt}\\
    \footnotesize (b) Lognormal SVD $\chi=2$ MPS 
   \end{tabular} \qquad
  \begin{tabular}[b]{c} 
    \includegraphics[width=.31\linewidth,trim={0 0 0cm 0},clip]{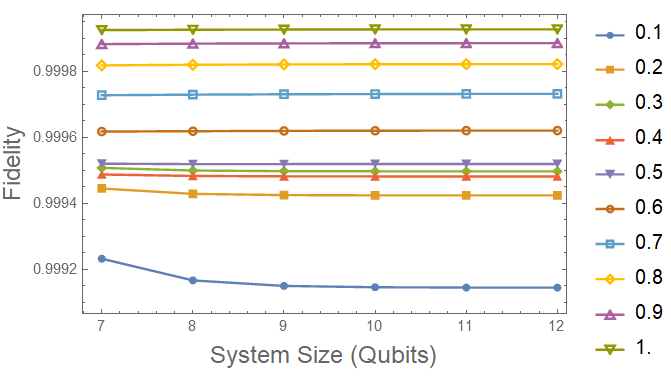}
        \label{fig:lorentzian_opt}\\
    \footnotesize (c) Lorentzian SVD $\chi=2$ MPS
    \end{tabular}
  \begin{tabular}[b]{c}
    \includegraphics[width=.25\linewidth,trim={0 0 0cm 0},clip]{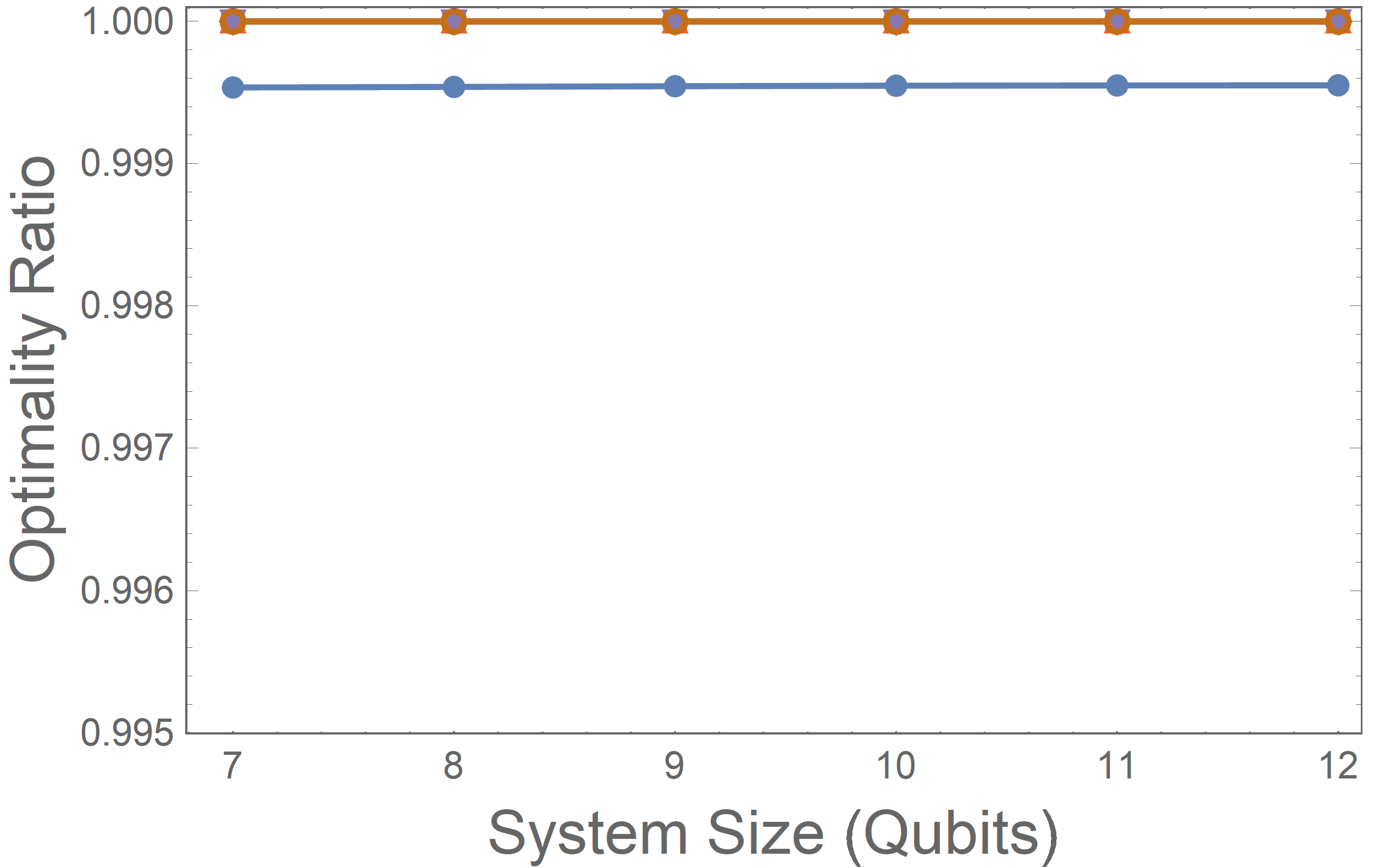} 
        \label{fig:gaussian_ratio}\\
    \footnotesize (d) Gaussian Optimality Ratio
  \end{tabular} \qquad
  \begin{tabular}[b]{c} 
    \includegraphics[width=.25\linewidth,trim={0 0 0 0},clip]{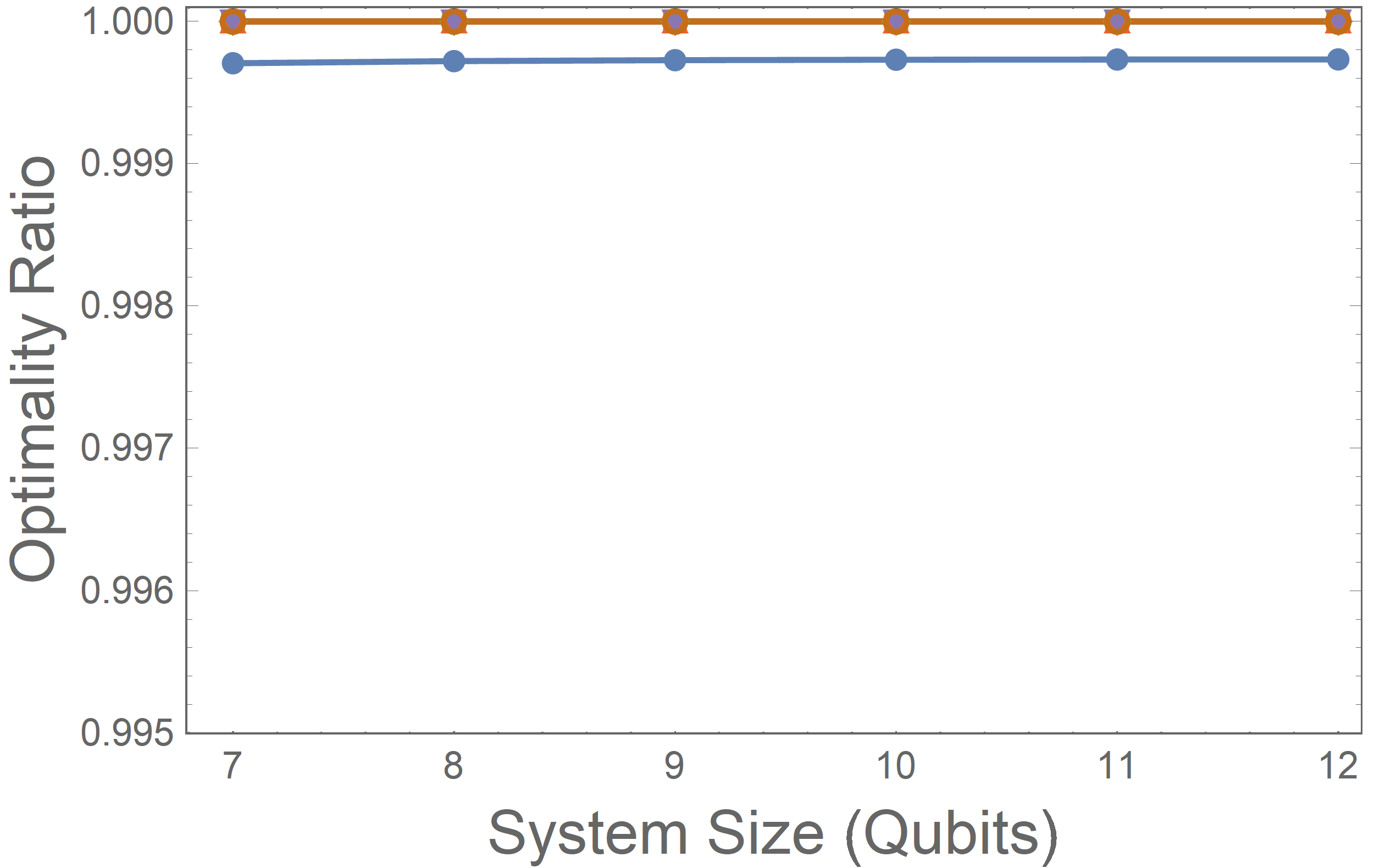}
        \label{fig:lognormal_ratio}\\
    \footnotesize (e) Lognormal Optimality Ratio
   \end{tabular} \qquad
  \begin{tabular}[b]{c} 
    \includegraphics[width=.32\linewidth,trim={0 0 0cm 0},clip]{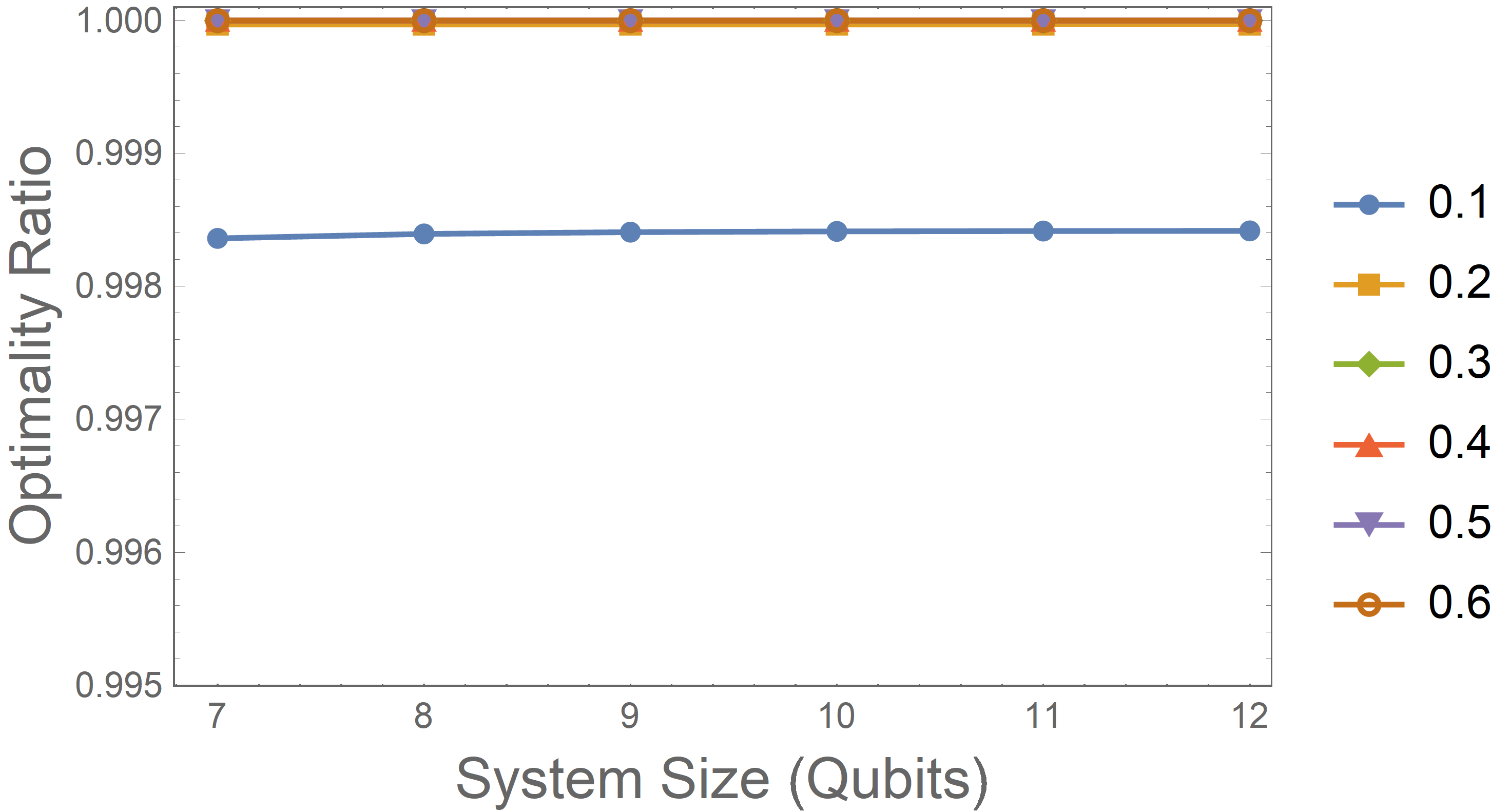}
        \label{fig:lorentzian_ratio}\\
    \footnotesize (f) Lorentzian Optimality Ratio
    \end{tabular}
  \caption{(a-c) Scaling of SVD $\chi=2$ MPS fidelities, as generated by the exact Algorithm \ref{alg:mps_svd} for (a) Gaussian, (b) lognormal, and (c) Lorentzian distributions. (d-f) Scaling of the optimality ratio (\ref{eq:optimality_ratio}) of the optimal low-rank MPS as generated by the exact Algorithm \ref{alg:mps_svd} against the MPS generated by Algorithm \ref{alg:main_algorithm} for (d) Gaussian, (e) lognormal, and (f) Lorentzian distributions.}
  \label{fig:opt_ratio_mps_analysis}
    \vspace{-0.5cm}
\end{figure*}
To evaluate the accuracy of our technique, for footnotesize systems we explicitly compare the constructed state with the target. This is done with the \textit{fidelity} measure $\mathcal{F}$ \cite{nielsen2002quantum}, defined as:
\begin{align}
    \mathcal{F}(\ket{\phi},\ket{\psi}) &= \text{Tr}\sqrt{\rho^{1/2} \sigma \rho^{1/2}}\
\end{align}
where $\rho = \ket{\phi}\bra{\phi}$ and $\sigma = \ket{\psi}\bra{\psi}$.
Using this measure, we can estimate how exactly the constructed states match the targets.

Figure \ref{fig:results_analysis} depicts the fidelities of circuits generated by Algorithm \ref{alg:main_algorithm} plotted against the standard deviation of the targeted SDR distribution, in different regimes. We expect to see that as $\sigma$ approaches 0 fidelity should decrease as the MPS is no longer able to accurately capture the large maximum upper bound on the derivative of the distribution. In Figure \ref{fig:results_analysis} this occurs for $\sigma \sim 0.12$, closely matching predictions by the spectral analytical modeling in equation (\ref{eq:exp_decay_bounds}). This can be attributed to equation (\ref{eq:sdr_entropy}), where the added entropy of the state grows with the maximum derivative of the SDR function $\Tilde{f}'$. For very squeezed states, this large effective constant dominates equation (\ref{eq:sdr_entropy}), and we find low-$\chi$ MPS states unable to accurately represent the function. Additionally, we find a region $\sigma \approx 0.3$ in which accuracy slightly decays as well. This reflects a fluctuation of the rank of the distributions in this region, beneath the upper bounds set by the derivatives. It is also indicative of error caused by the fixed $k=3 \implies 8$ region subdivision.

\subsection{Error Analysis}

Error arises in three places within Algorithm \ref{alg:main_algorithm}: the piecewise polynomial approximation $\Tilde{g}(x)$, the compression of the total MPS, and gate extraction. Empirically, we find that the dominant error comes from \textbf{MPS compression}, followed by the approximation error of the piecewise polynomial function (PP error). Gate extraction error is negligable. Figure \ref{fig:stacked_error} displays the decomposition of the total error in each constructed distribution and attributes each to its source. Among all constructed cases, MPS error contributes to $99.9\%$ on average, while PP and gate extraction error account for $0.11\%$ and $0.01\%$ on average, respectively. For the lognormal distribution, PP error contributes more significantly than in any other distribution, accounting for $0.92\%$ for $N = 7$ qubit constructions.

 The polynomial function approximation error can be decreased by increasing the degree of the fit polynomial. Doing so comes at the cost of increasing constants in the computational complexity of the entire procedure, and for large values relative to the system size the practical implementation complexity may begin to be affected. Figure \ref{fig:degree_study} displays sensitivity of Algorithm \ref{alg:main_algorithm} to the order of polynomial approximator used, studied for a single instantiation of each SDR function: $\sigma = 0.1$. Accuracy increases monotonically for increasing approximation degree, with diminishing returns beginning at the second order. Even for this difficult set of squeezed functions, cubic polynomials are able to construct the states with over $99\%$ accuracy, and increasing to $5^{\text{th}}$ order polynomials increases the fidelity up to $99.8\%, 99.91\%,$ and $99.95\%$ for each of the Gaussian, lognormal, and Lorentzian distributions, respectively. 

\subsection{Optimality Ratios}

The scalability of the technique rests on the conjecture that $\chi=2$ MPS representations remain good approximators for the target functions as the system size scales up. Empirically this can be estimated by tracing the fidelity of the SVD $\chi=2$ MPS representations to the SDR function and scaling up system size. The SVD $\chi=2$ MPS can be determined with Algorithm \ref{alg:mps_svd}, which does not include any form of function approximation as a component. Figure \ref{fig:opt_ratio_mps_analysis} shows these curves for a selection of standard deviations, and presents numerical support for the claim that these $\chi=2$ approximations remain consistently accurate for larger and larger systems. As a result, the gate extraction component of Algorithm \ref{alg:main_algorithm} remains unaffected by an increase in the size of the system overall.

The main factor in the accuracy of Algorithm \ref{alg:main_algorithm} is then the construction of the approximate $\chi=2$ MPS. In order to estimate how accurately the constructed state matches the optimal, we can use an \textit{optimality ratio} metric: 
\begin{align}
    \mathcal{R} = \frac{\mathcal{F}_{\text{circuit}}}{\mathcal{F}_{\text{optimal}}}
    \label{eq:optimality_ratio}
\end{align}

For more squeezed states the absolute fidelity of the state constructed by Algorithm \ref{alg:main_algorithm} remains high and the optimality gap remains constant or slightly improves. This is strong evidence that accuracy will scale well with larger system discretizations.

\section{Applications}
\label{sec:applications}
A primary application for this procedure is any Monte Carlo style quantum algorithm that estimates the value of an observable evaluated on classical probability distributions. Monte Carlo methods have been shown to have quadratic speedup in these domains \cite{montanaro2015quantum}, and to demonstrate this we discuss the Amplitude Estimation based financial derivatives pricing algorithm. 

Many algorithms have been proposed in the risk analysis and financial derivatives pricing context \cite{stamatopoulos2019option,woerner2019quantum,rebentrost2018quantum,rebentrost2018mc,orus2019quantum,ramos2019quantum}. Many of these are based around the same principle: after solving a variation of the Black-Scholes equation that dictates the movement of an asset, encode this solution as a probability distribution into a quantum state. Once this is complete, then a linear function is computed on this state, which represents a superposition over all values of this linear function, weighted by probability. This computes an expected value of this operator, which in many cases is formulated specifically to encode the price of a financial derivative of the asset. The expected value is evaluated using Amplitude Estimation \cite{brassard2002quantum}, with error and convergence that scales as:
\begin{align*}
    T = \mathcal{O}(T_{\text{prep}}/\varepsilon_{\text{sampling}})
\end{align*}
Where $\varepsilon$ is the sampling error. Classical Monte Carlo techniques for the same function scale as $\mathcal{O}(1/\varepsilon^2)$, giving the quantum algorithm a quadratic speedup. Clearly though, there is dependence on $T_{\text{prep}}$, or the time required to encode this distribution. With Algorithm \ref{alg:main_algorithm}, this can be done in linear time with tuneable accuracy.
\subsection{Extensions and Future Work}
One benefit of this work is that it extends to any real-valued functions that are well-approximated by a set of piecewise low-order polynomials. This work thus extends to cover classical input data sets, so long as the data can be well approximated with an analytical generating function. One example of this is image data, which often can be approximated or interpolated into an approximate analytical form \cite{keys1981cubic,lee1997scattered,hou1978cubic,parker1983comparison}. This in theory would allow for quantum states corresponding to image data to be efficiently constructed, given a method for multivariate MPS encoding.

\section{Conclusion}
\label{sec:conclusion}
In this work we develop an algorithm to prepare quantum states corresponding to smooth, differentiable, real-valued functions. The algorithm constructs a linear-depth circuit of arbitrary two-qubit quantum gates, and does so only requiring linear computation time. Evaluating the accuracy of this technique empirically on commonly used probability distributions shows that high degrees of accuracy are able to be obtained even for the most squeezed target functions. These techniques require computation that scales linearly with system size, and there is no evidence that accuracy decays as systems increase in size, showing promise for the scaling of these techniques to much larger systems. 



\bibliographystyle{plain}
\bibliography{references}

\end{document}